# Fast-speed and low-power-consumption optical phased array based on thin-film lithium niobate platform

Zhizhang Wang,[†] Xueyun Li,[†] Jitao Ji, Zhenxing Sun, Jiacheng Sun, Bin Fang, Jun Lu, Xiangfei Chen, Shining Zhu, and Tao Li[*]

National Laboratory of Solid State Microstructures, Key Laboratory of Intelligent Optical Sensing and Manipulations, Jiangsu Key Laboratory of Artificial Functional Materials, School of Physics, College of Engineering and Applied Sciences, Nanjing University, Nanjing 210093, China

**Abstract**

Fast scanning-speed and low-power-consumption are becoming progressively more and more important in realizing high-performance chiplet optical phased arrays (OPAs). Here, we establish an integrated OPA based on thin-film lithium niobate-on-insulator (LNOI) platform to access these outstanding performances. Significantly, a lithium niobate (LN) OPA chip is implemented by 32/48 channels LN waveguides enabled by electro-optic modulations, which showcases the low power consumption (1.11nJ/$\pi$) and fast operation speed (14.4 ns) promising the advantage of the LNOI platform for integrated OPAs. As results, we experimentally achieved a beam steering with a 62.2°×8.8° field of view (FOV) and a beam divergence of 2.4°×1.2°. Moreover, by employing sparse aperiodic arrays in waveguides design we obtained a significant reduction of lateral divergence to 0.33° for the radiation beam. This work demonstrate that remarkable advantage of LNOI platform for power-saving and scalable OPA chips for various applications.

## 1. Introduction

Optical phased arrays (OPAs) have gained significant attention for applications in areas of light detection and ranging (LiDAR) [1-3], free-space communication [4, 5], imaging projection [6, 7], and remote sensing [8]. In recently years, with the bloom of autonomous driving, OPAs have been envisaged as a promising beam steering solution for an all-solid-state LiDAR due to its high integration, compactness and scalability [9, 10]. In particular, silicon (Si) platforms have made significant success in integrated

OPAs because of its mature fabrication and compatibility with complementary metal-oxide semiconductor (CMOS) processes [11-15]. However, the two-photon absorption, high third-order nonlinearity and large power-consumption in silicon further limit the scalability and output beam power of OPAs based on Si platform [16, 17]. Silicon nitride ($Si_3N_4$) has been highlighted as an alternative platform to overcome those disadvantages by virtue of low nonlinearity and propagation loss [18, 19], however the scalability and power consumption of $Si_3N_4$ OPA has been also limited by the lower thermo-optical coefficient of $Si_3N_4$. So, the scheme of hybrid $Si/Si_3N_4$ integration have been proposed, but it will increase the complexity of fabrication process and systems [20-23].

Lithium niobate (LN) exhibits a wide transparent window (0.35-5μm), low absorption loss, piezoelectric effect, high second-order nonlinear optical coefficient, and relatively high Pockels electro-optic coefficient [24-26]. Especially, the electro-optic (EO) property of LN is very suitable for the phase modulation in OPA, which provide a high modulation efficiency and low optical losses at the same time. As earlier as 1974, OPAs based on the waveguides of epitaxial LN film have been reported by Tien et al. [27], achieving the beam deflection and the scan angle reach to 4°. However, traditional bulk LN waveguides are fabricated mostly by metal diffusion, ion implantation and proton exchange, which leads to large bent radii and impedes its large-scale photonic circuits. With the breakthroughs of the thin-film LN-on-insulator (LNOI) in fabrication techniques over the last few years, the LNOI has become a high-performance integrated photonics platform which combines the superior optical properties and EO effect of LN material with large refractive index contrast of LNOI, making large-scale integrated OPAs possible in the future [28-32].

Here, we demonstrate a monolithically integrated LN-OPA that reduces the power consumption while maintaining high operation speeds for integrated photonics chip. Using the LNOI platform, the power consumption of an EO phase shifter is reduced to 1.11 nJ per π phase shift and response time is cut down to 14.4 ns. As a proof of concept, a 32-channel array manages to sweep a 62.2°×8.8° field of view (FOV) in two dimensions by using EO modulation and wavelength tuning, and beam divergence of 2.4°×1.2° at λ=1550 nm. In addition, a 48-channel with sparse aperiodic arrays are used to decrease the lateral divergence of output beam with the angle reduces to 0.33°. Our work demonstrates LN-OPA as a scalable and power-efficient solution for various applications in the future, including all-solid-state LiDAR, free-space communication, etc.

## 2. Results and discussion

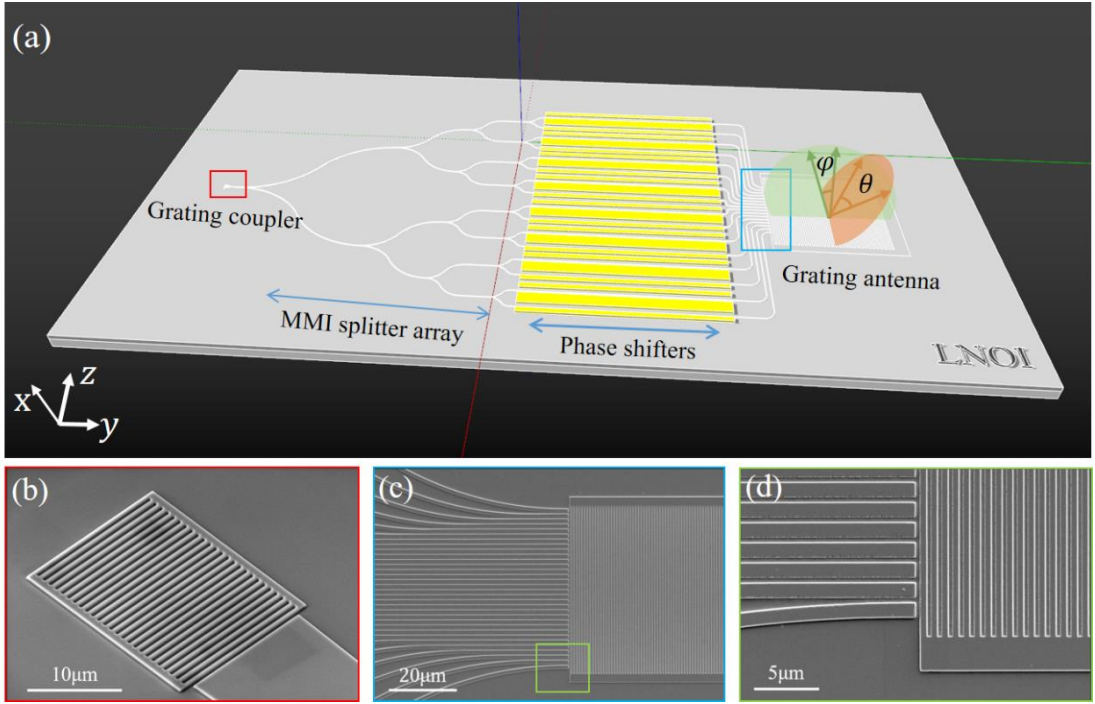

Figure 1 OPA chip based on LNOI. (a) Schematic illustration of the OPA for 2D beam steering. Scanning-electron microscope (SEM) images of the grating coupler and waveguide arrays are shown in (b), (c), and a close-up of the waveguide array in (d).

Figure 1(a) shows the schematic of the LN-OPA chip, which consists of a grating coupler, cascaded multi-mode interference (MMI) trees, EO phase shifters, and grating radiation antennas. Here, the OPA chip was fabricated in an X-cut LNOI platform (NANOLN) comprising a 600-nm-thick top LN film and buried oxide layer with a thickness of 2 μm. Ridge waveguides are designed 1 μm width with a 300 nm slab thickness and 300 nm ridge height (see Supplementary 1, Fig. S1). Firstly, the light around 1550 nm was controlled by polarization controller (PC) for maintaining TE-polarization, and coupled into the chip through the grating coupler (Fig. 1(b)), and then was split into 32 channels by a cascaded MMI splitter tree. Each channel passed through an EO phase shifter, which employs travelling-wave electrode design with length of 8 mm and the 4.5 μm pitch. At the end of the phase shifter array, the spacing of adjacent waveguides quickly decrease to wavelength (1.5 μm) by bending the waveguide, as shown in Fig. 1(c) and 1(d), and the crosstalk among waveguides is negligible.

Different from conventional waveguide grating antennas, a slab grating as a single radiation antenna with an etched thickness similar to waveguides has been used for increase the scanning range and simple fabrication process. The beam interferes in the LN slab and radiate out to space through the grating, as shown in Fig. 1(d). In order to describe the emission angle of an OPA, beam steering angles have been defined in the lateral ($\theta$ axis) and longitudinal ($\varphi$ axis) directions by EO phase modulation and operating wavelength tuning, respectively. Depending on the linear phase difference $\Delta\phi$ and the grating equation, the lateral and longitudinal sweeping angle $\theta$ and $\varphi$ are determined by

$$\sin\theta = \frac{\lambda_0 \Delta\phi}{2\pi d}, \tag{1}$$

$$\sin\varphi = \frac{\Lambda n_{eff} - \lambda_0}{n_0 \Lambda}, \tag{2}$$

where $\Delta\phi$ is the phase difference of adjacent waveguides, $d$ is the spacing of waveguide array, $n_0$ is the refractive index of the air, $\Lambda$ is the slab rating period, $n_{eff}$ is the effective index of the grating, and $\lambda_0$ is the wavelength of the beam in the free space. When the spacing larger than about half of a wavelength, the OPA with uniformly spaced emitters generate grating lobes that can be expressed as $\pm\sin^{-1}(\frac{\lambda}{2d})$. For the slab grating antenna, the antenna spacing is 1.5 μm, so theoretically a scanning range greater than ±31° can be achieved.

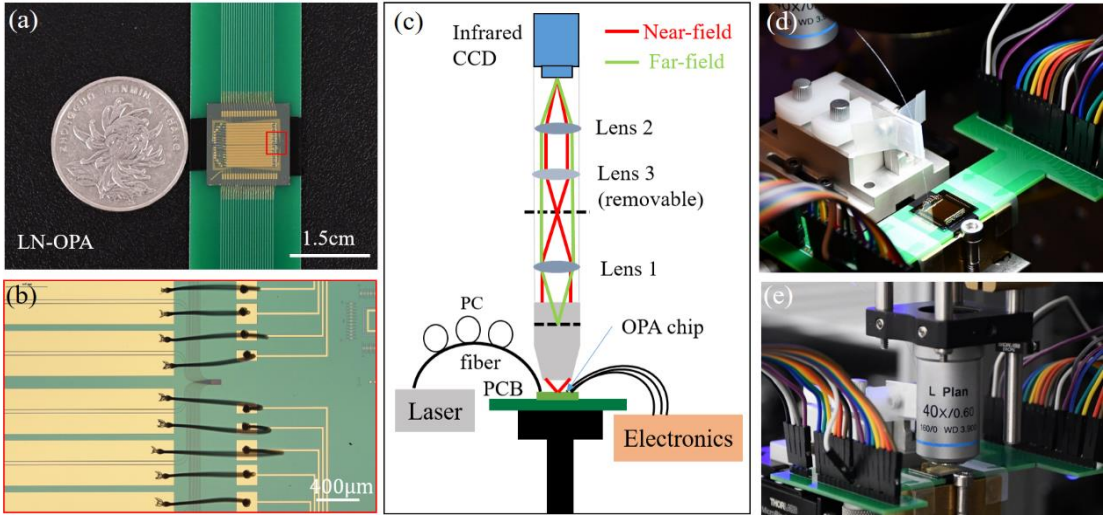

Figure 2 (a) Photograph of the 1.5 cm × 1.5 cm LN-OPA chip wire-bonded to a printed circuit board. (b) Microscopic photograph of wired-bonded electrodes and the region of grating antennas indicated by red box in (a). (c) Schematic of the near/far-field measurement setup. (d), (e) Photographs of fiber-to-grating coupling and testing in the experiment.

As shown in Fig. 2(a), a 32-channel LN-OPA is fabricated on a 1.5 cm × 1.5 cm X-cut LNOI chip, which is wire-bonded on a printed circuit board (PCB) and controlled by digital-to-analog converters (DAC). Figure 2(b) shows the detailed microscopic photograph of wired-bonded electrodes and the region of grating antennas. Due to a total phased array aperture of 48 μm×100 μm, we adopted an objective lens with a high NA (0.6) to achieve higher performance far-field images and designed the emitting angle near 0° in the longitudinal direction at an input wavelength of 1550 nm. As shown in Fig.2 (c), a near/far-field measurement setup placed vertically was used to characterize the OPA beam steering, which brings convenience of switching between near-field and far-field images by employing/removing Lens 3. The near-field image (both lenses; red rays) is used to align the imaging system to the region of emitter antennas, and then the Fourier image (Lens 1 and 2; green rays) correspond to the far-field ones. Figure 2(d) and 2(e) show photographs of the fiber-to-grating coupling and testing in the experiment, respectively. The infrared camera capturing the far-field image and DAC were connected to a computer simultaneously, and the output electric signals will be changed by analyzing the far-field image. With the captured images as feedback, a particle swarm optimization (PSO) algorithm was employed to calibrate the initial phase and form a well beam spot in the far field (see Supplementary 1, Section 2).

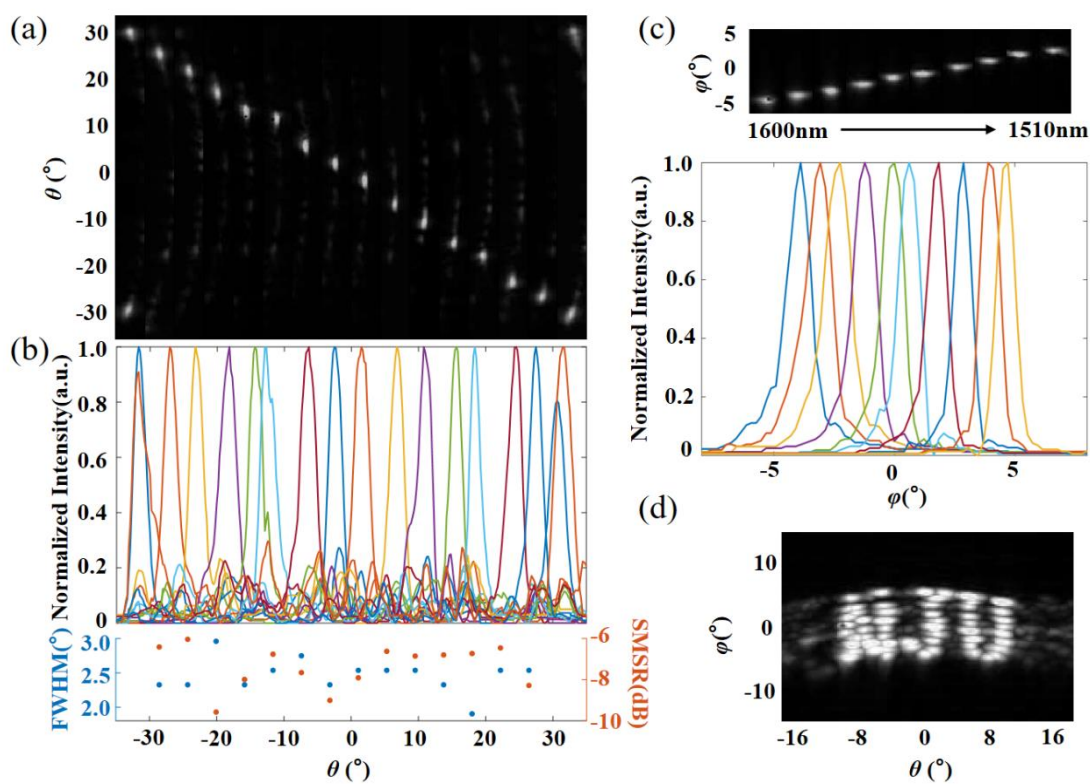

Figure 3 Characterization of far-field emissions of the 32-channel LN-OPA. (a) The spliced image of the far-field radiation pattern with the beam scanning range of ±31.1° in $\theta$ direction. (b)Measured normalized far-field intensity along the $\theta$ direction. Bottom inset: full width at half-maximum (FWHM) and side

mode suppression ratio (SMSR) of each radiative beam with an average FWHM of 2.4° and the average SMSR of -7.4 dB (see Visualization 1 showing the bean steering in $\theta$). (c) Beam steering in $\varphi$ direction and corresponding normalized intensity distribution by tuning the wavelength from 1510 nm to 1600 nm.(d) Normalized far-field beam steering with wavelength tuning and EO modulation, where a 'NJU' pattern is formed covering a FOV of 23.2°×8.8°.

As shown in Fig. 3(a) and 3(b), we experimentally achieve aliasing-free beam steering from -31.1° to 31.1° at a fixed wavelength (1550 nm) in lateral direction, and the average side mode suppression ratio (SMSR) is <-7.4 dB when the beam steered within ±31.1°range. As plotted in the bottom part of Fig. 3(b), the steering beams are characterized to be with an average full width at half-maximum (FWHM) of 2.4°, which is mainly limited by the size of the radiation aperture (48 μm) in lateral direction. By sweeping the laser wavelength from 1510 nm to 1600 nm, the emission light can achieve a FOV of 8.8° and a FWHM of 1.2° in longitudinal direction (see Fig. 3(c)). The divergence in the wavelength-steered axis can be further improved by employing a well-controlled shallow grating etch. So far, the capability of lateral and horizontal scanning of OPA has been individually demonstrated with a FOV of 62.2°×8.8° and a FWHM beam size of 2.4° ×1.2°. In addition, we utilize the wavelength tuning to slice the 2D pattern and perform lateral scanning through EO modulation to condense three letters of "NJU" with a FOV of 23.2°×8.8° as shown in Fig. 3(d), which verifies the capability of the LN-OPA for two-dimensional beam scanning. It is a non-ignorable phenomenon that the far field beam spot has a process of unstable state when the modulation voltage applied, which may be impeded by the photorefractive (PR) effect inherent in LN. We also found that the far field beam spot can be a stabilized and repeatable state when gradually increase the scan rate (the influence of PR effect can be reduced by removing the dielectric cladding [33]). In fact, our results have confirmed that the PR effect is effectively mitigated when the scan rate is higher than the response time of PR effect (see more discussions in Supplement 1, Section 3).

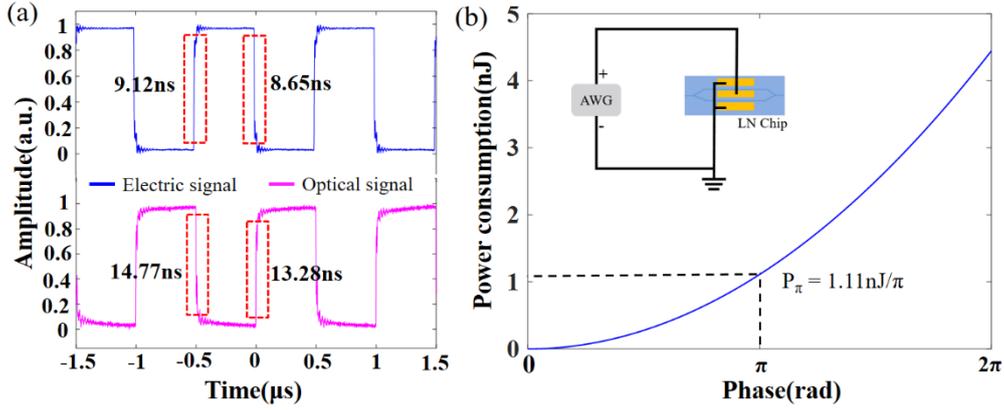

Figure 4 (a) Measured response time when applying a square voltage signal, indicating a switching speed of about 14 ns. (b) The power consumption of a phase shifter as a function of phase difference between two arms of Mach-Zehnder interferometer, inset illustrates the equivalent circuit of the phase shifter which can be seen as a capacitor.

Compared with conventional OPAs based on other material platforms, the LN-OPA possess significant advantages of operation speed and power consumption owing to the excellent EO effect of LN material. For the power consumption and operation speed of the phase shifter, the rising and falling of LNOI Mach-Zehnder interferometer (MZI) phase shifter has been tested to reflect the power consumption and response time of the OPA. A 1 MHz square voltage signal is applied on the phase shifter and the modulated light is transformed into electric signals by a photodetector (PD) connected with an oscilloscope (see Supplementary 1, Fig. S4). Figure 4(a) illustrates the waveforms of input electrical signal and response optical signal and indicates the rising and falling time up to 13.28 ns and 14.77 ns, which exhibits extremely fast optical response time. Furthermore, as shown in the inset of Fig. 4(b), the MZI phase shifter is considered to be equivalent to a capacitor and its capacitance is measured to be 39.5pF by an Inductance-Capacitance-Resistance (LCR) digital bridge tester. With the measured half-wave voltage ($\pi$ phase shift at 7.5 V by each phase shifter ) $V_\pi$ of 7.5 V, the power consumption as a function of the induced phase shifts between two arms of MZI is plotted in Fig. 4(b) and estimated to be 1.11 nJ/$\pi$ per channel. It is worth noting that the energy stored in the MZI is almost consumed by peripheral circuits and thus the actual power consumption of the LN-OPA should be much lower than that estimated above (see Supplementary 1, Section 3 and 4).

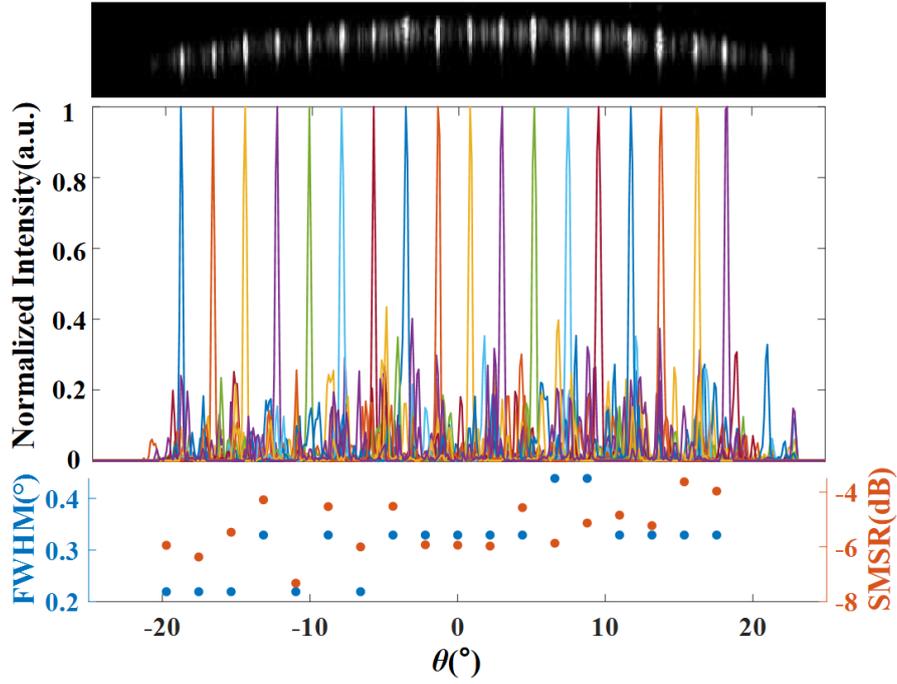

Figure 5 Characterization of far-field emissions of aperiodic spacing antennas of 48-channel LN-OPA. Top: the sliced profile of far-field pattern. Middle: corresponding normalized intensity distribution as a function of $\theta$. Bottom: FWHM and SMSR of each radiative beam, showing a FWHM beam size of 0.33° and SMSR of about 5 dB.

In order to achieve smaller beam divergence in $\theta$, we extend the phase channels from 32 to 48 and adopt a sparse aperiodic arrangement of arrays to achieve a larger radiation aperture, and a genetic algorithm is used to iterate the position of each channel with SMSR as the target function. As demonstrated above, the divergence angle in $\theta$ is limited by number of phase channels and dense waveguide arrangement. As a result, the optimized 48-channel LN-OPA with an average pitch of 5.6 $\lambda$ and the total aperture size of 410 μm is fabricated and experimentally investigated, with a result of the FWHM of 0.216° and SMSR of 11 dB for beam steering of 0° in theory (see Supplementary 1, Section 5). As shown in Fig. 5, the scanning FOV of the LN-OPA achieves ±20° and the FWHM beam size is effectively narrowed down to about 0.33° with SMSR of about 5 dB. Although the optimization for FWHM results in a slightly decrease in FOV and SMSR, these two performances could be improved by increasing the number of phase channels. With the sparse aperiodic arrangement of waveguide channels, the FWHM is decreased to one-sixth compared with the uniform spacing antennas of 32-channel OPA, providing a feasible solution to decrease the lateral divergence of output beam and enhance the performance of the OPA.

Table 1 The comparison of performance metrics of OPAs based on different platforms

| Platform | Principle | Modulation power | Speed | FOV (a°×b°)(SMSR) | FWHM (a°×b°) | Channels | Ref. |
|---|---|---|---|---|---|---|---|
| SOI | Plasma dispersion effect | 2 μW/π | 30 μs | ±28×15(12 dB) | 0.04 | 512 | [21] |
| SOI | Thermo-optics effect | 22 mW/π | - | ±33×3.3(7 dB) | - | 24 | [11] |
| $Si_3N_4$-Si | Thermo-optics effect | 20 mW/π | 22.8 μs | ±48×14 | 2.3×2.8 | 32 | [22] |
| $Si_3N_4$-Si | Plasma dispersion effect | 1.8 μW/π | 30 μs | ±70×19.4(7.4 dB) | 0.021×0.1 | 128 | [23] |
| SOI | Thermo-optics effect | 7 mW/π | - | ±70×13.5(13.2 dB) | 2.1×0.08 | 64 | [12] |
| **This Work** | **Pockels EO effect** | **1.11 nJ/π*** | **14.4 ns** | **±31.1×8.8(7.4 dB)** <br> **±20×8.8(5 dB)** | **2.4×1.2** <br> **0.33×1.8** | **32** <br> **48** | |

★ **For Pockels EO effect, the power consumption is description by energy change.**

For the performance of the proposed LN-OPA, Table 1 shows a comparison of performance metrics of the OPAs based on different platforms. Conventional OPAs have been predominantly restricted by operation speed (around μs) and excessive power consumption to modulate the phase. In this work, by leveraging the exceptional electro-optic properties of LN, the proposed LN-OPA can operate at ultra-high speed, almost three orders of magnitude over other OPAs. Besides, the power consumption of the LN-OPA is comparable to that of the OPAs based on plasma dispersion effect (which however suffers from high optical loss) and far outperforms the OPAs based on thermo-optics effect. The LNOI platform provides a compact, low-power-consumption and scalable to implement high-performance integrated OPA chips. Although the FOV and divergence of OPAs just have conventional performances, it should be mentioned that the LN-OPA is chosen to be with 32/48 channels for a proof of concept in this work, the FOV, SMSR and FWHM of the OPA could be further improved by employing other optimization methods in future. More importantly, owing to the incomparable advantages in operation speed and power consumption, the LN-OPA is of potential scalability and higher performances.

## 3. Conclusions

In summary, we have demonstrated a high-efficient optical phased arrays based on LNOI platform, which decreases the power consumption to the level of nJ per π phase shift and improves the operation speed to nanosecond simultaneously. As a proof of concept, the LN-OPA chip has achieved beam steering of a 62.2°×8.8° FOV with the beam divergence of 2.4°×1.2°. By using LNOI platform, the scalability of OPAs limited by the high power consumption of their embedded large number of phase shifter would be remarkably improved. With the advantage of scalability and low power consumption, the LNOI photonics platform facilitates the prospect of a power-efficient, compact, and large-scale integrated OPA for a wide applications including long-range LiDAR, autonomous vehicles and free-space communications.

**Funding.** National Key Research and Development Program of China (2022YFA1404301), National Natural Science Foundation of China (12174186, 92250304, 62288101).

**Acknowledgment.** The authors acknowledge the Micro-fabrication Center of the National Laboratory of Solid State Microstructures (NLSSM) for technique support. Tao Li thanks the support from Dengfeng Project B of Nanjing University. The authors thank NanZhi Institute Advanced Optoelectronic Integration for OPA chip fabrication.

**Disclosures.** The authors declare no conflicts of interest.

**Data availability.** Data underlying the results presented in this paper are not publicly available at this time but may be obtained from the authors upon reasonable request.

**Supplemental document.** See Supplement 1 for supporting content.

**Author contribution:** T. L. and Z. W. conceived the idea. Z. W. proposed the design and performed the numerical simulations; J. S. fabricated the samples; X. L. developed the algorithm to calibrate the initial phase; Z. W., X. L., J. J., Z. S., and J. L. performed the optical measurements; Z. W. and T. L. analyzed the results and wrote the manuscript with the input from all authors; T. L. supervised the project.

†These authors contributed equally to this work.

# Fast-speed and low-power-consumption optical phased array based on thin-film lithium niobate platform: Supplementary material

Zhizhang Wang,† Xueyun Li,†Jitao Ji, Zhenxing Sun, Jiacheng Sun, Bin Fang, Jun Lu, Xiangfei Chen, Shining Zhu, and Tao Li*

National Laboratory of Solid State Microstructures, Key Laboratory of Intelligent Optical Sensing and Manipulations, Jiangsu Key Laboratory of Artificial Functional Materials, School of Physics, College of Engineering and Applied Sciences, Nanjing University, Nanjing 210093, China

## 1. Fabrication process of LN-OPA

A commercially available X-cut lithium niobate-on-insulator (LNOI) wafer (NANOLN) with a 600nm thick lithium niobate (LN) layer and a 2 μm buried silicon dioxide is first cleaned and a layer of ma-N2405 is spin-coated on the LN thin film as a mask. The waveguide patterns are defined by an E-beam lithography (EBL) process on the resist and transferred 300 nm deep into LN with an optimized argon plasma and $CHF_3$ in an inductively coupled plasma (ICP) etching system subsequently. After removing the resist and cleaning the LN sidewall, two layers of LOR5B film and AZ5214 film are spin-coated and followed by an ultraviolet (UV) exposure to define the electrode patterns. After development, a 400 nm thick gold film is deposited onto the wafer and then a lifted-off process transfers the patterns. Finally, the fabricated chip is mounted onto a printed circuit board (PCB) with two-step wire-bonding which connects the electrodes of phase shifters to the surrounding pads and the pads to the PCB, respectively (Fig. S1).

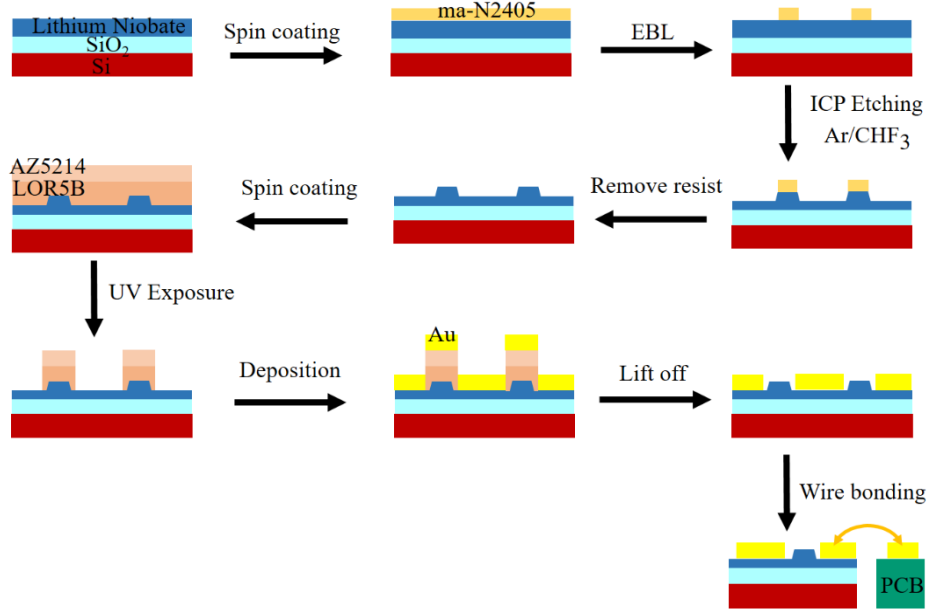

Figure S1 The fabrication procedure of the proposed LN-OPA.

## 2. Far field radiation calibration

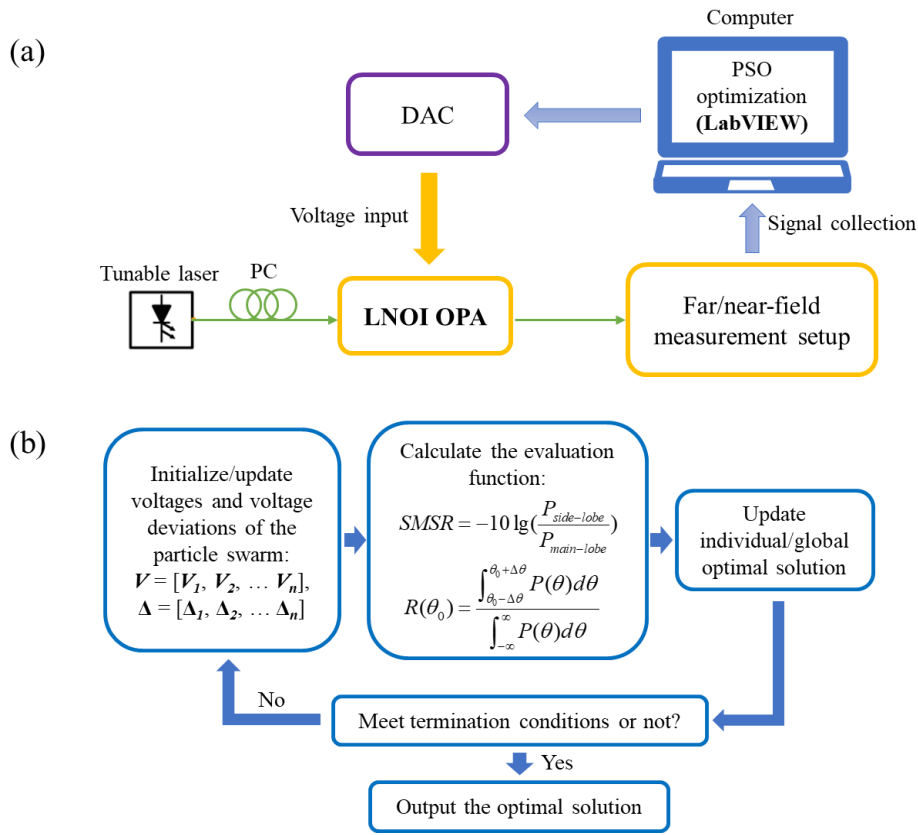

Figure S2 (a) The testing system for phase calibration, utilizing a particle swarm optimization (PSO)

algorithm implemented in NI LabVIEW code. (b) The flowchart of PSO algorithm. DAC: digital – analog converter. PC: polarization controller.

In the experiment, due to the phase distortion caused by fabrication imperfection such as rough sidewall of LN waveguides and processing deviation of bending waveguides, phase calibration at the emitter is necessary to suppress side lobes and improve steering efficiency. Figure S2(a) illustrates the testing system for the phase calibration based on particle swarm optimization (PSO) algorithm implemented in NI LabVIEW code. The optimization is executed on a computer connected with the near infrared charge coupled device (CCD, RAPTOR/OW1.7-CL-640) and the digital-to-analog converter (DAC, NI9264), acquiring far-field images and updating electric voltages to the LN-OPA respectively.

The PSO algorithm starts with a set of arbitrary locations (corresponding to voltages applied to the phase shifters) and a set of arbitrary velocities (corresponding to voltage deviations) which compose the property of particle swarm (see Fig. S2(b)). The far-field images after Fourier transformation by far/near-field measurement setup are taken and evaluated for their side mode suppression ratio (SMSR) and central energy ratio $R(\theta_0)$ as the composite evaluation function which gives the criterion to update the individual (global) optimal solution in the optimization process. The updated velocity set and the location set in the next iteration are calculated by

$$\Delta_{i+1} = w \times \Delta_i + c_{self} \times r_1 \times (p_{best} - V_i) + c_{social} \times r_2 \times (g_{best} - V_i), \tag{S1}$$

$$V_{i+1} = V_i + \Delta_{i+1}, \tag{S2}$$

where $V_i$ and $\Delta_i$ is the location set and the velocity set in the $i$-th iteration, $w$ is the predefined inertia weight, $c_{self,\ social}$ is the predefined individual/social factor, $r_{1,2}$ is the random number between 0 and 1, $p_{best}$ and $g_{best}$ are the individual and global optimal solution respectively. When the termination conditions are satisfied, the optimization program will be stopped, and corrected far-field beam spot at the expected steering angle is acquired when the optimal voltage array applied.

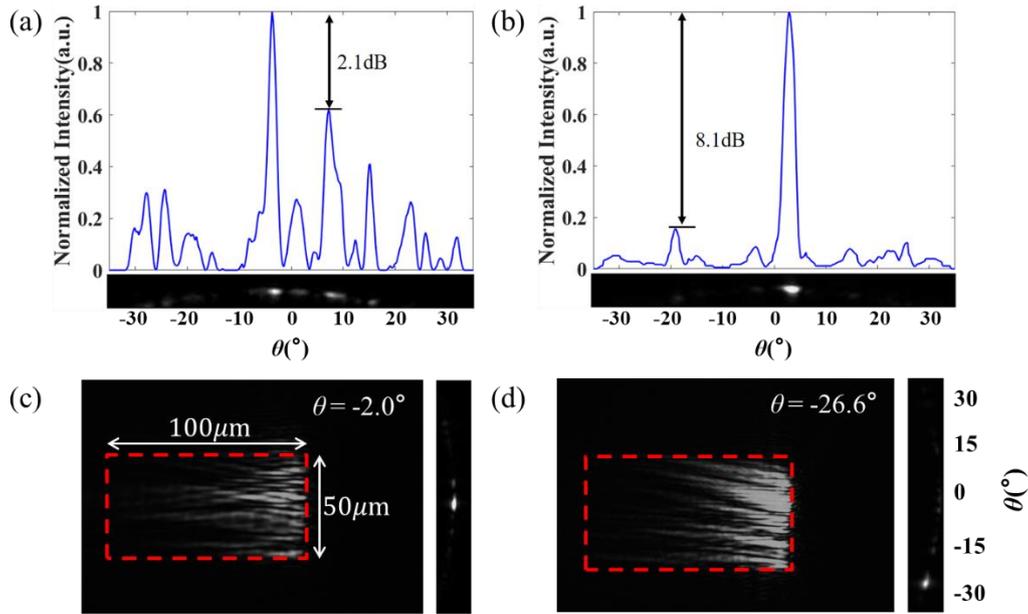

Figure S3 (a), (b) The far-field intensity distribution before and after calibration respectively. (c), (d) Near/far-field images with steering angle of -2.0° and -26.6° after phase calibration. Red dashed line indicates the area of grating antennas of 48 μm × 100 μm size.

Due to the near/far-field measurement system is established, the phase of LN-OPA is calibrated for desired steering angle and validated its performance. Figure S3(a) and S3(b) present the comparison of far-field intensity before and after calibration, showing an improvement from 2.1 dB to 8.1 dB. After calibration process, the near-field and far-filed images are respectively captured. As shown in Fig. S3(c) and S3(d), optical field distribution in the near-field has tilted wavefront at the different angles of 2.0° and -26.6°, which corresponds to the measured beam deflection angles in the far field.

## 3. The measurement of response time

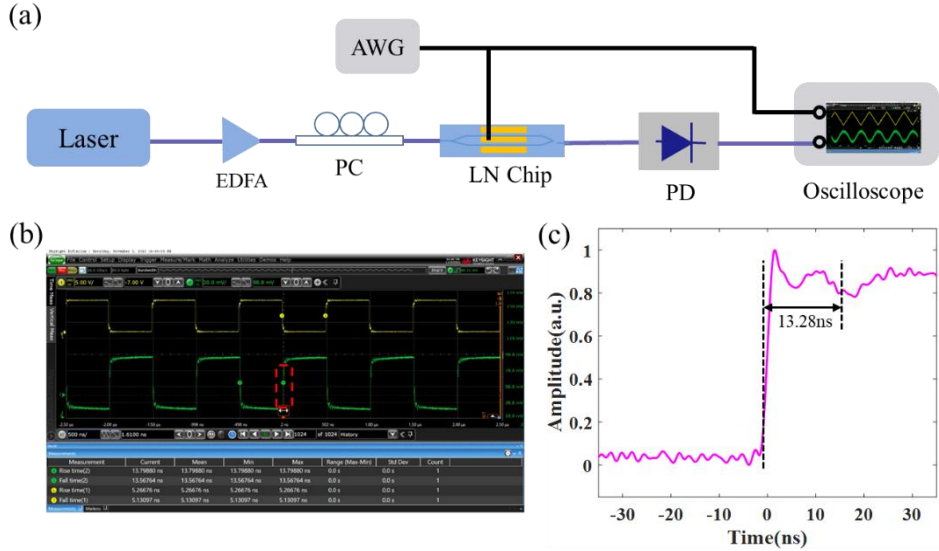

Figure S4 (a) Experimental setup for response time and half-wave voltage measurement. EDFA: erbium-doped fiber amplifier. PC: polarization controller. AWG: arbitrary waveform generator. PD: photodetector. (b) Oscilloscope screenshot of the response time. (c) Magnified view of a rising edge, indicating the switching time of 13.28 ns.

Regarding the speed of the OPA, we tested the rising and falling edges of the Mach-Zehnder interferometer (MZI) phase shifter. The experimental setup for response time measurement is presented in Fig. S4(a). Light is emitted from a tunable laser (Keysight 8164B), amplified by an erbium-doped fiber amplifier (EDFA) and then passes through a polarization controller (PC) before being injected into the chip. An arbitrary waveform generator (AWG, Tektronix AFG3251) controls the electric voltage on the electrodes of the MZI phase shifter and modulates the light passing through the chip which is detected by a photodetector (PD) and an oscilloscope (Keysight MXR054A). With a 1 MHz square waveform signal of half-wave peak to peak voltage applied on the chip, the average rising time of 13.28 ns and falling time of 14.77 ns are achieved as presented in Fig. 4(a) and Fig. S4(b) and the magnified view of a rising edge is exhibited in Fig. S4(c).

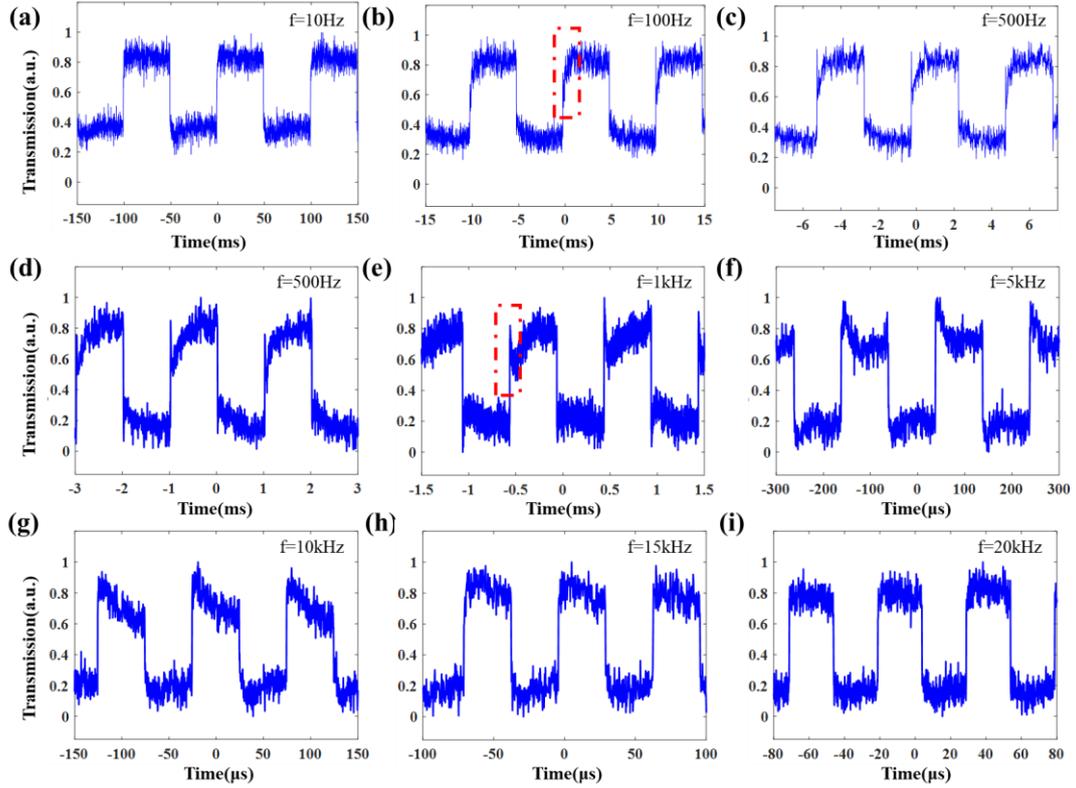

Figure S5. Voltage drifts caused by photorefractive effect. Rectangle waveform signals of different frequency from 10 Hz to 20 kHz and identical peak-to-peak voltage is applied on the phase shifter, corresponding optical response shown in (a)-(i) respectively.

As an inherent property of LN, the photorefractive (PR) effect impedes its usage in practical application and appears in our experiment of OPA testing. The PR effect induces the drift of far-field beam spot, which is mitigated by increasing the scanning rate. To further investigate the relation between the drift and the scanning rate for drift-free operation, we applied a series of square waveform signals of identical peak-to-peak voltage and various frequency from 10 Hz to 20 kHz on the LN MZI phase shifter and detected the optical outputs. Results show that when the voltage changes, the power of transmitted light presents a slow rise of several milliseconds (see red dashed box in Fig. S5(b)) after a rapid peak within about 50 μs (see red dashed box in Fig. S5(e)). With the frequency increasing, the drift is gradually mitigated and vanished at the frequency of 20 kHz, where the impact of PR effect can be effectively eliminated.

## 4. The principle and measurement of power consumption

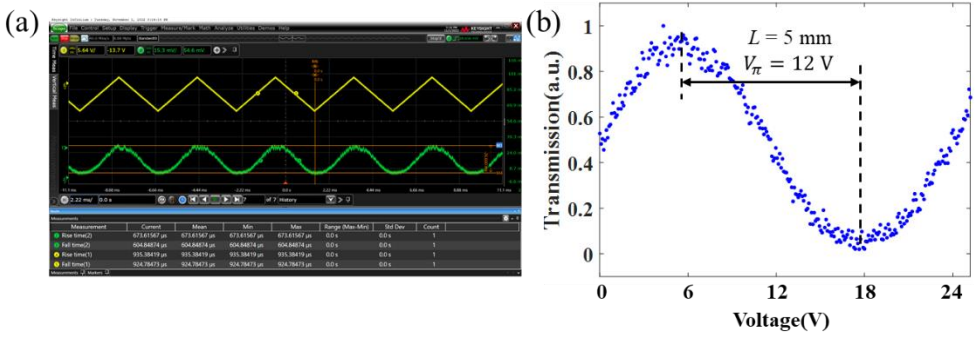

Figure S6 (a) Oscilloscope screenshot of the half wave voltage measurement. (b) Measured normalized optical transmission of a 5 mm phase shifter, indicating the $V_\pi$ value of 12 V.

For traditional OPAs based on thermo-optics effect, the power consumption is predominantly from the heat dissipation. By contrast, the phase shifter of LN-OPA is based on Pockels electro-optics effect and equivalent to a capacitor, which have inherently no energy dissipation but energy change. The energy changed in the phase shifter can be calculated by

$$E = \frac{1}{2}CV_{rms}^2, \quad (S3)$$

where $C$ is the capacitance and $V_{rms}$ is the root-mean-square voltage. With the measurement setup illustrated in Fig. S4(a), we experimentally acquired the half wave voltage of 12 V for a $L$=5 mm phase shifter, implying the voltage-length product of 6 V•cm (shown in Fig. S6(a) and S6(b)), and correspondingly the half wave voltage of the $L$=8 mm LN MZI phase shifter which has the same configuration as LN-OPA phase shifters is 7.5 V. For the capacitance, an Inductance-Capacitance-Resistance (LCR) digital bridge tester (Victor VC4090A) is utilized and the average value of the 8 mm length LN phase shifter is measured to be 39.5 pF. Consequently, the power consumption is calculated to be 1.11 nJ/π by Equation S3.

## 5. The optimization of aperiodic spacing antennas

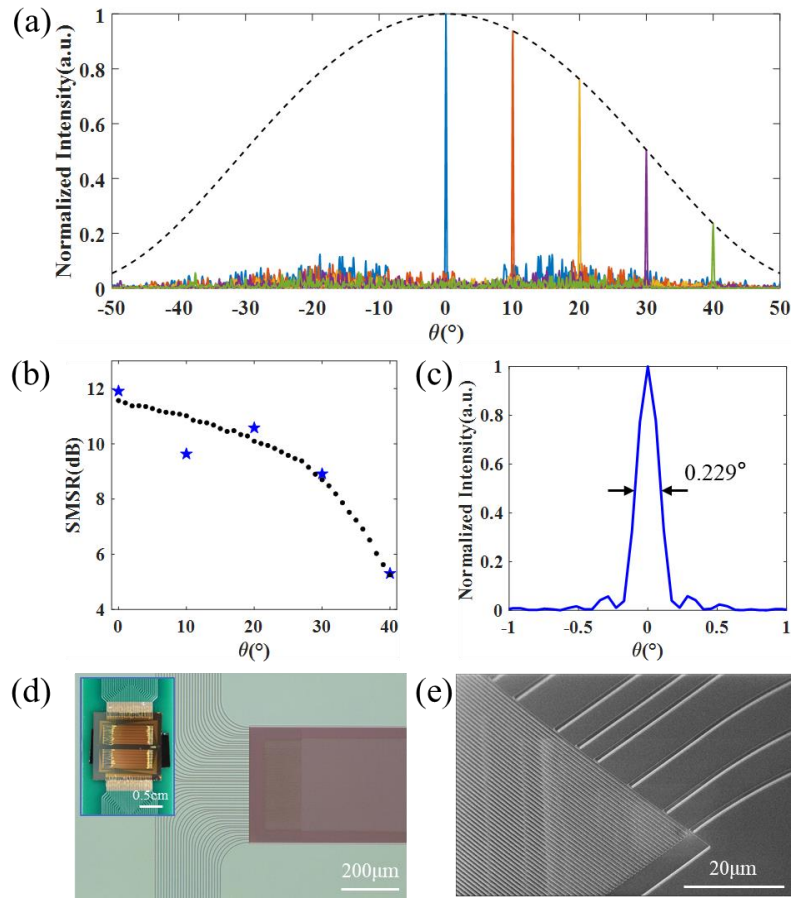

Figure S7 (a) Simulated far field intensity distribution at different steering angles of designed aperiodic antennas. (b) The side mode suppression ratio (SMSR) as a function of steering angle of calculated (black dashed line) and simulated results (blue five-pointed star). (c) Magnified view of the simulated far field beam at 0° with a full width at half-maximum (FWHM) of 0.229°. (d) Microscopy image of the emitting grating of the fabricated LN-OPA chip based on the proposed aperiodic spacing antennas design. Inset: holistic photograph of the LN-OPA chip. (e) Scanning-electron microscope (SEM) image of part of the antennas.

In the application, the beam divergence is a vital metric associated with detection range and precision and expected to be as small as possible. The direct approach to decrease the divergence angle is enlarging the radiative aperture size, such as adding channels and increasing the pitch between adjacent waveguides. Limited by footprint of the chip and the fabrication complexity, number of channels cannot dramatically rise up, and therefore the sparse aperiodic in waveguides design is the feasible way for OPAs. We utilize a sparse aperiodic spacing design to suppress grating lobes and a genetic algorithm (GA) is applied for a higher SMSR by optimizing the position of each channel with an average pitch in the range of 3λ to 8λ. The optimized 48-channel waveguide array has a size of 410 μm in the direction of phase modulation and an average pitch of 5.6λ. Simulation results show that the aperiodic antennas is capable to steer beam within ±30° FOV maintaining a SMSR more than 8 dB and the FWHM at the 0° emitting is 0.229° (see Fig. S7(a-c)). As the proof of concept, we fabricated the LN-OPA with the sparse aperiodic spacing antennas design and experimentally tested its performance. Figure S7(d) and S7(e) show the microscopy image of the emitting grating and the SEM photograph of the details at the antennas respectively with the full-view of the fabricated LN-OPA chip inset in Fig. S7(d).